\documentclass[a4paper,11pt]{article}
\pdfoutput=1
\usepackage{jcappub}
\usepackage[T1]{fontenc}

\usepackage[normalem]{ulem}
\usepackage{cleveref}
\usepackage[sort&compress]{natbib} 

\usepackage{graphicx,xcolor}
\definecolor{CiteBlue}{RGB}{45,52,151}
\hypersetup{
    colorlinks=true,
    linkcolor=CiteBlue,
    urlcolor=CiteBlue,
    citecolor=CiteBlue
}

\usepackage{amssymb}
\usepackage{amsmath}
\usepackage{booktabs}

\usepackage{siunitx}
\DeclareSIUnit{\parsec}{pc}

\usepackage{bm}
\newcommand{\bb}[1]{\bm{\mathrm{#1}}}
\newcommand{\du}{\mathrm d}
\newcommand{\refcite}[1]{Ref.~\cite{#1}}
\newcommand{\refscite}[1]{Refs.~\cite{#1}}
\newcommand{\dm}{\chi}
\newcommand{\dr}{\eta}
\newcommand{\kd}{\mathrm{kd}}
\newcommand{\kr}{\mathrm{kr}}
\newcommand{\eq}{\mathrm{eq}}
\newcommand{\fs}{\mathrm{fs}}
\newcommand{\class}{\textsc{class}}
\newcommand{\kdc}{\dot{\kappa}_{\dm}}
\newcommand{\kddr}{\dot{\kappa}_{\dr\dm}}
\newcommand{\hMpc}{\,h\,\SI{}{\mega\parsec^{-1}}}
\newcommand{\dbar}{\bar{\du}}

\title{\boldmath Kinetic recoupling of dark matter}

\author[a]{Benjamin V. Lehmann,}
\author[b,c]{Logan Morrison,}
\author[b,c]{Stefano Profumo,}
\author[b,c]{and Nolan Smyth}

\affiliation[a]{\ignorespaces
	Center for Theoretical Physics,
	Massachusetts Institute of Technology,\\
	Cambridge, MA 02139, USA
}

\affiliation[b]{\ignorespaces
	Department of Physics, University of California, Santa Cruz\\
	Santa Cruz, CA 95064, USA
}
\affiliation[c]{\ignorespaces
	Santa Cruz Institute for Particle Physics,\\
	Santa Cruz, CA 95064, USA
}

\emailAdd{benvlehmann@gmail.com}
\emailAdd{loanmorr@ucsc.edu}
\emailAdd{profumo@ucsc.edu}
\emailAdd{nwsmyth@ucsc.edu}

\abstract{
	We study the possibility that dark matter re-enters kinetic equilibrium with a radiation bath after kinetic decoupling, a scenario we dub \emph{kinetic recoupling}. This naturally occurs, for instance, with certain types of resonantly-enhanced interactions, or as the result of a phase transition. While late kinetic decoupling damps structure on small scales below a cutoff, kinetic recoupling produces more complex changes in the power spectrum that depend on the nature and extent of the recoupling period. We explore the features that kinetic recoupling imprints upon the matter power spectrum, and discuss how such features can be traced to dark matter microphysics with future observations.
}

\begin{document}
\maketitle

\flushbottom

\section{Introduction}
\label{sec:introduction}

While the identity of dark matter (DM) is still unknown, its cosmological role
has come into sharp definition. Standard $\Lambda$CDM cosmology accurately predicts the structure and evolution of the Universe on large scales. The great success of such a simple theory is one of the crowning triumphs of the particle DM paradigm. However, the success of $\Lambda$CDM cosmology is a source of frustration for the study of DM particle physics, since large-scale observables are readily matched by an extremely broad space of microphysical models. Regardless of the fundamental nature of DM, a dark sector is generally compatible with observational data as long as the DM is produced cold, attains the appropriate abundance, and has at most feeble nongravitational interactions.

The universality seen in predictions for large-scale observables becomes much more fragile on small scales. Thus, a key strategy to distinguish between microphysical models is to study the distribution of matter at smaller and smaller length scales. Small-scale cosmological observables have long been a source of intriguing opportunities for DM model building. Indeed, the $\Lambda$CDM paradigm has sometimes been said to face a ``small-scale crisis'' \cite{Weinberg:2013aya,DelPopolo:2016emo,Bullock:2017xww,Perivolaropoulos:2021jda}: the missing-satellites problem \cite{Moore:1999nt,Klypin:1999uc,Strigari:2007ma,Diemand:2007qr}, the too-big-to-fail problem \cite{Boylan-Kolchin:2011lmk,Boylan-Kolchin:2011qkt,Garrison-Kimmel:2014vqa,Jiang:2015vra}, the core-cusp problem \cite{Moore:1999gc,deBlok:2009sp,Amorisco:2011hb}, and other small-scale anomalies have pointed to particle physics beyond the minimal $\Lambda$CDM picture. While many of the tensions may be accounted for by baryonic physics or artifacts in the predictions \cite{Bullock:2000wn,Benson:2001at,Kazantzidis:2003hb,Kravtsov:2004cm,Simon:2007dq,Brooks:2012ah,Garrison-Kimmel:2013yys,Kim:2017iwr,Verbeke:2017rfd,Ostriker:2019}, these small-scale tensions continue to motivate extensive work on structure formation in nonminimal scenarios, including DM self-interactions \cite{Spergel:1999mh,vandenAarssen:2012vpm,Kamada:2013sh,Kaplinghat:2015aga,Tulin:2017ara,Binder:2017lkj,Yang:2021kdf} and modified thermal histories \cite{Colin:2000dn,Sigurdson:2003vy,Kaplinghat:2005sy,Cembranos:2005us,Boyarsky:2008xj,Viel:2013fqw,Anderhalden:2012qt,Boehm:2000gq,Green:2003un,Boehm:2004th,Green:2005fa,Bertschinger:2006nq,Bringmann:2009vf,Cornell:2013rza,Cyr-Racine:2013fsa,Visinelli:2015eka,Bringmann:2016ilk}. More recently, the $H_0$ and $S_8$ tensions \cite{DiValentino:2016hlg,DiValentino:2017gzb,Vagnozzi:2019ezj,Lin:2019htv,Verde:2019ivm,Allahverdi:2020bys,       Ishak:2018his,Benisty:2020kdt} have presented an independent hint that $\Lambda$CDM cosmology is incomplete, and have revitalized the study of models with features such as nonminimal dark energy \cite{Zhao:2017cud,Lambiase:2018ows,Keeley:2019esp,Niedermann:2020dwg,Gomez-Valent:2020mqn,Lucca:2021dxo,DiValentino:2021izs} and nontrivial thermal histories \cite{Pandey:2019plg,Heimersheim:2020aoc,FrancoAbellan:2020xnr,Joseph:2022jsf}. Such additional ingredients typically modify structure formation as well, so refined predictions and observations of small-scale structure promise to offer a key probe of dark sectors in the coming decades. 

The study of small-scale structure is challenging both theoretically and observationally. Nonlinearity presents a key challenge: as density perturbations grow, their evolution enters the nonlinear regime, and standard analytical predictions become unreliable. Small-scale modes enter the nonlinear regime earlier than large-scale modes, and modes at wavenumbers $k \gtrsim 0.1\hMpc$ are evolving nonlinearly in the late Universe. Traditionally, studying nonlinear structure formation has required expensive $N$-body simulations. Still, motivated by the rich physics opportunities, a pair of substantial programs in theory and observation promises to make significant progress at small scales in the coming years. (See e.g. the recent reports in \refscite{Chang:2022lrw, Annis:2022xgg}).

From the perspective of DM model building, the first feature that might emerge from these programs is a suppression of structure on small scales. This occurs in several classes of nonminimal DM models. For example, a well-known possibility is that the DM is not quite ``cold'', but rather undergoes significant free-streaming upon decoupling from the thermal bath. This ``warm DM'' scenario \cite{Maccio:2009isa,Drewes:2016upu} erases density fluctuations smaller than the free-streaming scale. A different possibility is that the DM remains in kinetic equilibrium with a relativistic thermal bath to late times, well beyond the typical decoupling temperature of weakly interacting massive particles (WIMPs). This late kinetic decoupling produces acoustic oscillations in the DM power spectrum, featuring a prominent suppression of power at small scales \cite{Boehm:2000gq,Hofmann:2001bi,Chen:2001jz,Green:2005fa,Bertschinger:2006nq,Bringmann:2009vf,Gondolo:2012vh,Cornell:2013rza,Dasgupta:2013zpn,Bringmann:2016ilk,Gondolo:2016mrz,Ko:2016uft,Dror:2017gjq, Profumo:2006bv,Boehm:2004th}. This has been confirmed in dedicated cosmological simulations \cite{Vogelsberger:2015gpr,Schewtschenko:2015rno}, and has led to a large body of theoretical work on the front of DM model building \cite{Boehm:2000gq,Boehm:2001hm,Chen:2001jz,Hooper:2007tu,vandenAarssen:2012vpm,Boehm:2014vja,Bringmann:2016ilk}.

Predictions for the small-scale matter distribution in scenarios like these have been systematized for a broad class of models in the context of effective field theory (EFT) \cite{Cornell:2013rza}. Microphysical models that admit an EFT description can be linked directly to the power spectrum via frameworks such as the \textbf{E}ffective \textbf{Th}eory \textbf{o}f \textbf{S}tructure (ETHOS) \cite{Vogelsberger:2015gpr,Cyr-Racine:2015ihg,Munoz:2020mue}. However, models which cannot be described by an EFT are not encapsulated by such frameworks, and the potential impact of such interactions on the matter distribution is not well understood. The systematic study of such models is challenging. Still, it is possible to identify scenarios that should not be accommodated in simple EFTs, and to study their phenomenological consequences.

That possibility motivates the present work. We study the simplest and most generic scenario that does not readily admit an EFT description, and thus presents novel predictions for small-scale observables. Specifically, we consider an alternative to late kinetic decoupling: after decoupling from a relativistic thermal bath, the DM may subsequently \emph{re-enter} thermal equilibrium with said bath, a phenomenon we dub \emph{kinetic recoupling}. First, we investigate the conditions under which this can occur within the realm of standard cosmology. Second, we demonstrate how such a kinetic recoupling would manifest itself in observations. Finally, we clarify how kinetic recoupling could actually be discovered from upcoming small-scale cosmological observations, and how such observations can directly inform DM model building.

This paper is organized as follows. In \cref{sec:decoupling}, we review the significance of kinetic decoupling to structure formation, and discuss the different mechanisms that affect structure on small scales. In \cref{sec:recoupling}, we examine the possibility of a period of kinetic recoupling, introduce concrete realizations of such a scenario, and discuss the effects on structure formation. In \cref{sec:numerics}, we numerically study the impact of recoupling on the matter power spectrum, and discuss observational signatures. We conclude with a discussion of the implications of our results in \cref{sec:discussion}.

We note that the term ``kinetic recoupling'' was first introduced by \refcite{Kamada:2016qjo} in the context of Coulomb scattering of charged massive particles (CHAMPs) with baryons. This scenario differs from the one studied in this work in that the CHAMPs recouple to nonrelativistic charged particles rather than to a relativistic radiation bath. We compare these cases further in \cref{sec:recoupling}.

\subsection*{Notation and conventions}
Throughout this paper, we use $\dm$ to denote the DM species and $\dr$ to denote the interacting radiation species. To simplify phase space integrals, we write $\dbar^n\bb p \equiv \du^n\bb p/[(2\pi)^n\,E]$. We write $h = H_0 / (\SI{100}{\kilo\meter/\second/\mega\parsec})$, where $H_0$ is the Hubble constant today. Dots ($\dot a$) denote differentiation with respect to conformal time. When writing cosmological perturbations, we work in conformal Newtonian gauge.

\section{Decoupling and the matter power spectrum}
\label{sec:decoupling}
Our study of kinetic recoupling is motivated in part by the well known effects of a late kinetic decoupling. Thus, before introducing the recoupling scenario, we review the role of kinetic decoupling in structure formation, and the impact of delayed decoupling on the matter distribution. The results we quote here are discussed at length in \refscite{Boehm:2000gq,Green:2003un,Boehm:2004th,Green:2005fa,Bertschinger:2006nq,Bringmann:2009vf,Cornell:2013rza,Cyr-Racine:2013fsa,Visinelli:2015eka,Bringmann:2016ilk}.

Note that we are interested in studying the final kinetic decoupling between DM and \textit{any} radiation bath. It is not necessary that the radiation bath involved be a part of the visible sector. To make our results as simple and as general as possible, we will assume that the DM $\dm$ is maintained in kinetic equilibrium by scattering with a single radiation species $\dr$. We impose no particular assumptions on the nature of this radiation species in most of our results, and it could in principle be a Standard Model field. However, our fiducial case of interest is one in which the radiation species is another member of the dark sector, so we will refer to $\dr$ as ``dark radiation'' or ``DR'' for brevity. In general, additional relativistic degrees of freedom in equilibrium at the same temperature as the Standard Model bath are strongly constrained by $\Delta N_{\mathrm{eff}}$ \cite{Jedamzik:2009uy,Boehm:2013jpa,Calabrese:2011hg}, so it is also useful to allow the dark sector to have a different temperature from the visible sector in order to relax these bounds. We thus introduce an initial temperature ratio $\xi \equiv T_\dr/T_\gamma$ and allow $\xi < 1$.

\subsection{Thermal dark matter and kinetic decoupling}
Before discussing the details of late kinetic decoupling, we review the definition of kinetic equilibrium and kinetic decoupling, and distinguish between chemical and kinetic decoupling. To that end, we return briefly to the basic formalism for the evolution of the dark matter distribution. In this section, we write $p = \left|\bb p\right|$ and denote 4-momenta with capital letters ($P$).

Heuristically, the DM is said to be in kinetic equilibrium with another species when the rate of scattering processes is high enough to maintain a common temperature between the two baths. This is opposed to chemical equilibrium, in which the rate of number-changing processes is high enough to maintain an equilibrium distribution of particle number between the two baths, so that the number density of each species is given by the equipartition theorem. Kinetic equilibrium can be maintained in the absence of number-changing processes, and it generally persists for some time after chemical equilibrium is lost. Indeed, since chemical equilibrium is generally maintained by annihilation processes (e.g. $\dm\dm\leftrightarrow\dr\dr$) that exchange energy between the two baths, chemical equilibrium typically implies kinetic equilibrium.

Chemical and kinetic equilibrium are ultimately statements about the DM and DR phase space densities, $f_\dm(\bb p)$ and $f_\dr(\bb p)$. In turn, the phase space density is governed by the Boltzmann equation \citep{Arcadi:2011ev,Bi:2011qm,Fan:2014zua,Visinelli:2015eka,Diacoumis:2018nbq}. In an Friedmann–Lema\^itre–Robertson–Walker metric (FLRW) background, the Boltzmann equation takes the form
\begin{equation}
	E \bigl(
		\partial_t - H \bb p\cdot\nabla_{\bb p}
	\bigr) f_{\dm}(\bb p) = C[f_{\dm},\, f_\dr]
	,
\end{equation}
where $E$ is the DM energy, $\bb p$ is the DM 3-momentum, $H$ is the Hubble rate, and $C$ is the collision operator. The collision operator is separated into two terms: an annihilation term, $C_{\mathrm{ann}}$, and an elastic scattering term, $C_{\mathrm{el}}$.  The annihilation operator contains all number-changing processes. For DM annihilating into DR as $\dm(\bb{p}) + \bar{\dm}(\tilde{\bb{p}}) \leftrightarrow \dr(\bb{k}) + \bar\dr\bigl(\tilde{\bb{k}}\bigr)$, the annihilation term averaged over initial and final state spins is

\begin{multline}
	C_{\mathrm{ann}}[f_{\dm},\,f_\dr] = \frac{1}{8}\int
	\dbar^3\bb k\;
	\dbar^3\tilde{\bb p}\;
	\dbar^3\tilde{\bb k}\;
	\left(2\pi\right)^{4}\delta^{4}(P + \tilde{P} - K  - \tilde{K}) \\
	\times \bigg[
		\left|\mathcal{M}_{\dm + \bar{\dm} \to \dr + \bar\dr}\right|^{2}
		f_{\dm}(\bb{p})
		f_{\dm}(\tilde{\bb{p}})
		\left[1\pm f_\dr(\bb{k})\right]
		\bigl[1\pm f_\dr\bigl(\tilde{\bb{k}}\bigr)\bigr] \\
		- \left|\mathcal{M}_{\dr + \bar\dr \to \dm + \bar{\dm}}\right|^{2}
		f_\dr(\bb{k})
		f_\dr(\tilde{\bb{k}})
		\left[1\pm f_{\dm}(\bb{p})\right]
		\left[1\pm f_{\dm}(\tilde{\bb{p}})\right]
	\bigg]
	,
\end{multline}
where $+$ and $-$ correspond to Bose enhancement and Pauli blocking for bosons and fermions, respectively. On the other hand, for DM--DR scattering via $\dm(\bb{p}) + \eta(\bb{k}) \leftrightarrow \dm(\tilde{\bb{p}}) + \eta\bigl(\tilde{\bb{k}}\bigr)$, the elastic scattering term is
\begin{multline}
	C_{\mathrm{el}}[f_{\dm},\,f_\dr] = \frac{1}{8}\int
	\dbar^3\bb k\;
	\dbar^3\tilde{\bb p}\;
	\dbar^3\tilde{\bb k}\;
	\,\left(2\pi\right)^{4}\delta^{4}(p - \tilde{p} + k  - \tilde{k})
	\left|\mathcal{M}_{\dm + \gamma \leftrightarrow \dm + \gamma}\right|^{2} \\
	\times \left[
		f_{\dm}(\bb{p})
		f_\dr(\bb{k})
		\bigl[1\pm f_\eta\bigl(\tilde{\bb{k}}\bigr)\bigr]
		-
		f_{\dm}(\tilde{\bb{p}})
		f_\dr\bigl(\tilde{\bb{k}}\bigr)
		\left[1\pm f_\dr(\bb{k})\right]
	\right]
	,
\end{multline}
where we take $1 \pm f_\chi \approx 1$ since the DM is typically nonrelativistic when the scattering operator is relevant. When determining the DM relic abundance, in most cases it is valid to ignore the elastic term and consider the first moment of the Boltzmann equation, integrating against $g_{\dm}\int\dbar^3\bb p$, where $g_\dm$ denotes the number of internal degrees of freedom of $\dm$. This yields the standard evolution equation for the DM number density,
\begin{equation}
	\frac{\du n_{\dm}}{\du t} + 3Hn_{\dm} =
	-\left\langle\sigma v\right\rangle(n_{\dm}^{2} - n_{\dm,\eq}^{2}),
\end{equation}
where $n_{\dm} = g_{\dm}\int\dbar^3\bb p\,f_{\dm}(\bb p)$ is the actual DM number density, $n_{\dm,\eq}$ is the DM equilibrium number density, and $\left\langle\sigma v\right\rangle$ is the thermally-averaged annihilation cross section. As long as the annihilation rate $\Gamma_{\mathrm{ann}}\equiv\left\langle\sigma v\right\rangle n_{\dm}$ exceeds the Hubble rate, $n_\dm$ tracks $n_{\dm,\eq}$, and the DM is said to be in chemical equilibrium.

When the annihilation rate falls below the Hubble rate, $\Gamma_{\mathrm{ann}} \lesssim H$, the DM falls out of chemical equilibrium, freezing the comoving number density near its relic value. This is chemical decoupling. However, the DM and DR will continue to exchange momentum efficiently for a period of time. We can describe the dynamics during this phase by considering the second moment of the Boltzmann equation, dropping the annihilation term but retaining the elastic term. During this period, the DM is typically nonrelativistic, and the elastic term can be expanded in the limit of small momentum transfer per collision between DM and DR. The leading-order result is
\begin{equation}
	C_{\mathrm{el}}\left[f_\dm,\,f_\dr\right]
	\simeq
	\frac12\Gamma_{\mathrm{el}}(T)m_\dm\left[
		T m_{\dm} \frac{\partial^2}{\partial p^2}
		+\left(p + 2T\frac{m_{\dm}}{p}\right)
			\frac{\partial}{\partial p}
		+ 3
	\right]
	f_{\dm}(\bb{p}),
\end{equation}
where $T$ is the temperature of the DR bath and $\Gamma_{\mathrm{el}}(T)$ is the DM--DR scatting rate. Explicitly, the DM--DR scattering rate is
\begin{equation}
\label{eq:scattering_rate}
	\Gamma_{\mathrm{el}}(T) =
	-\frac{1}{48\pi^{3}g_{\dm}m_{\dm}^{3}}\int\du\omega\,
		\frac{\partial f_\dr}{\partial\omega}\,
	\omega^{4}
	\left\langle\left|\mathcal{M}\right|^{2}\right\rangle_{t},
	\qquad
	\left\langle\left|\mathcal{M}\right|^{2}\right\rangle_{t} =
	-\frac{1}{8\omega^4}
	\int_{-4k^{2}_{\mathrm{cm}}}^{0}
	\du t\,t\left|\mathcal{M}\right|^{2}
	,
\end{equation}
where $\omega$ is the DR energy. Here we have used the Mandelstam variable $t = (P-\tilde{P})^{2}$, the squared center-of-mass momentum $k_{\mathrm{cm}}^{2}$, and the relevant matrix element $\mathcal{M}$. Integrating the Boltzmann equation against $g_{\dm}\int\dbar^3\bb p\,p^{2}$, one obtains the differential equation
\begin{equation}
	\frac{\du T_{\dm}}{\du t} + 2 H T_{\dm} = \Gamma_{\rm{el}}(T) \left(T-T_{\dm}\right),
	\qquad
	T_{\dm}
	= \frac{1}{3}\left\langle\frac{p^{2}}{E}\right\rangle
	= \frac{g_{\dm}}{3n_{\dm}}\int\dbar^3{\bb{p}}\,p^{2} f_{\dm}(\bb{p})
	.
\end{equation}
Now $T_\dm$ can be interpreted as the DM temperature in a manner consistent with the equipartition theorem.

As with chemical decoupling, when the rate of momentum exchange becomes relatively small, interactions cease and the dark matter starts behaving as a pressureless fluid. This is kinetic decoupling. The quantitative condition for kinetic decoupling is slightly more complicated than that for chemical decoupling, since a typical scatter only changes the momentum of a DM particle by $\Delta p \equiv p_\dm / N \ll p_\dm$, where $N \simeq \sqrt{3/2}m_\dm/T \gg 1$ \cite{Hofmann:2001bi}. Thus, kinetic equilibrium is only maintained as long as the rate for $N$ scattering events to take place is faster than the expansion rate, that is, $\Gamma_{\mathrm{el}}/N \gtrsim H$. In the remaining sections, we work directly with the DM--DR drag opacity $\kddr$ ($\dot\kappa_{\textnormal{DR--DM}}$ in the notation of \refcite{Cyr-Racine:2015ihg}) instead of the scattering rate itself. Thus, we take $\kddr < \mathcal H$ to signify kinetic decoupling, where $\mathcal H$ is the conformal Hubble rate $\dot a/a$.

The timing of kinetic decoupling clearly influences the behavior of the DM phase space density $f_\dm$, and therefore the behavior of DM density perturbations. In the remainder of this work, we will be mainly interested in studying the matter power spectrum $P(k)$. For ease of comparison with the literature, we will generally compute the dimensionless matter power spectrum $\Delta_m^2(k) \equiv P(k) \frac{k^3}{2 \pi^2}$ in the following sections. Note also that all of the power spectra shown in this work are \emph{linear} power spectra evaluated at $z=0$. While this is appropriate for comparison with other works in the literature, such power spectra are nonphysical for wavenumbers in the nonlinear regime at $z=0$.

\subsection{Kinetic decoupling and structure formation}
\label{subsec:structure}

We now turn to the mechanisms by which a modified kinetic decoupling can affect the formation of structure on small scales. 
In the paradigm of standard cosmology, DM structure forms hierarchically: the smallest structures form at early times, and assemble into larger structures over cosmic time. The abundance of the smallest primordial halos thus determines the properties of small-scale structure in the late Universe. Small-scale structure is sensitive to damping processes that disrupt the formation of these initial halos. One such process is the free streaming of warm DM. If the DM species is initially in kinetic equilibrium and departs from equilibrium while relativistic, then DM particles will free-stream out of any local overdensities without scattering until they are cooled by cosmic expansion. The length scale traversed before the DM cools is the damping scale associated with warm DM. Any structures smaller than this scale will be effectively erased by free streaming. Thus, a warm DM species which decouples at a temperature $T \sim m_{\mathrm{WDM}}$ produces a cutoff in the matter power spectrum that can be constrained by late-time observables. This manifests as a cutoff in the halo mass function at a mass scale of \cite{Vogelsberger:2015gpr}
\begin{equation}
    M_{\mathrm{cut},\mathrm{WDM}} \simeq
    \SI{e11}{M_\odot} \left(
    	\frac{m_{\rm WDM}}{\SI{1}{\kilo\electronvolt}}
    \right)^{-4}\,h^{-1}
    .
\end{equation}
Conversely, the approximate cutoff corresponding to a cold species kinetically decoupling at $T=T_{\rm kd}$ reads
\begin{equation}
	M_{\mathrm{cut},\mathrm{kd}} \simeq
	\SI{5e10}{M_\odot} \left(
		\frac{T_\kd}{\SI{100}{\electronvolt}}\right)^{-3}\,h^{-1}.
\end{equation}

However, free streaming is not the only damping mechanism at early times. If the DM is in equilibrium with a radiation bath, \textit{collisional damping} suppresses the formation of structure below a particular length scale. Inhomogeneities in the radiation fluid are suppressed by free streaming, and the viscous coupling between the radiation and the DM fluid damps inhomogeneities in the latter. Collisional damping remains efficient until the DM falls out of kinetic equilibrium with the radiation bath, effectively preventing the formation of overdensities on scales below the horizon size. Thus, the longer the DM remains in equilibrium, the larger the associated collisional damping scale. The matter power spectrum is exponentially damped on a comoving scale
\begin{equation}
	k_\kd \approx
	1.8\left(\frac{m_\dm}{T_\kd}\right)^{1/2}
	\frac{a_\kd}{a_0}H_\kd,
\end{equation}
and thereafter, free streaming further damps the power spectrum on a scale
\begin{equation}
	k_{\fs} \approx
	\left(\frac{m_\dm}{T_\kd}\right)^{1/2}
	\frac{a_\eq/a_\kd}{\log\left(4a_\eq/a_\kd\right)}\frac{a_\eq}{a_0}H_\eq
	.
\end{equation}
Combining these two effects gives rise to a damping factor of the form
\begin{equation}
	\label{eq:kdamp}
	\left[1 - \frac23\left(\frac{k}{k_\fs}\right)^2\right]\exp\left[
		-\left(k/k_{\mathrm{damp}}\right)^2
	\right]
	, \qquad
	k_{\mathrm{damp}} = \left(\frac{1}{k_\fs^2}+\frac{1}{k_\kd^2}\right)^{-1/2}
	.
\end{equation}
Further, the matter power spectrum can acquire a dark-sector analogue of baryon acoustic oscillations, i.e., dark acoustic oscillations (DAO) \cite{Bertschinger:2006nq,Huo:2017vef}. Supposing that kinetic decoupling occurs at a conformal time $\tau_\kd$, the acoustic length of the DR at decoupling is $\ell\sim\tau_\kd/\sqrt3$, and DAO damps the spectrum differently in two regimes. First, for wavenumbers $10\tau^{-1} \lesssim k \lesssim \tau_\kd^{-1}$, acoustic oscillations of the DR are communicated to the DM gravitationally, leading to a smooth suppression averaged over these scales. The acoustic oscillations of the coupled DM--DR fluid itself introduce an exponential damping at a scale
\begin{equation}
	k_{\mathrm{DAO}} \sim
	\frac{1}{\tau_\kd}\left(\frac{T_\kd}{m_\dm}\right)^{-1/2}
		\left[\log\left(\frac{\tau_0}{\tau_\kd}\right)\right]^{-1}
	.
\end{equation}

A more rigorous treatment requires the analysis of the Boltzmann equations governing the evolution of perturbations in the DM fluid. Following the conventions used within the ETHOS framework \cite{Cyr-Racine:2015ihg}, the perturbations of the DM fluid are determined by two coupled differential equations
\begin{align}
	\label{eq:boltzmann-dm-1}
	&\dot\delta_\dm + \theta_\dm - 3\dot\phi = 0
	,\\
	\label{eq:boltzmann-dm-2}
	&\dot\theta_\dm - c_\dm^2k^2\delta_\dm + \mathcal H\theta_\dm - k^2\psi
		= \kdc \left(\theta_\dm - \theta_\dr\right)
	.
\end{align}
Here $\delta$ and $\theta$ are the density and velocity divergence perturbations, $\phi$ and $\psi$ are the gravitational potentials, $c_\dm$ is the DM sound speed, $\kdc$ is the DM drag opacity, and $\mathcal H$ is the conformal Hubble parameter. Only the isotropic perturbations $\delta_\dm$ and $\theta_\dm$ are nonnegligible for the nonrelativistic DM fluid. The DR fluid, on the other hand, can have large anisotropic perturbations with mutual dependence between all multipoles in an infinite hierarchy of Boltzmann equations. Following the notation of \refcite{Cyr-Racine:2015ihg}, we write the $\ell$th moment as $\Pi_{\dr,\ell}$, with $\delta_\dr \equiv \Pi_{\dr,0}$ and $\theta_\dr \equiv \frac34k\Pi_{\dr,1}$. The first two equations involve only $\delta_\dr$, $\theta_\dr$, and the shear stress $\sigma_\dr = \frac12\Pi_{\dr,2}$:
\begin{align}
	\label{eq:boltzmann-dr-1}
	&\dot\delta_\dr + \frac43\theta_\dr - 4\dot\phi = 0
	,\\
	\label{eq:boltzmann-dr-2}
	&\dot\theta_\dr + k^2\left(\sigma_\dr - \frac14\delta_\dr\right)- k^2\psi
		= \kddr \left(\theta_\dr - \theta_\dm\right)
	.
\end{align}
Here $\kddr$ is the DR drag opacity, written as $\dot\kappa_{\textnormal{DR--DM}}$ in the notation of \refcite{Cyr-Racine:2015ihg}. The remainder of the hierarchy takes the form
\begin{equation}
	\label{eq:boltzmann-dr-3}
	\dot\Pi_{\dr,\ell} + \frac{k}{2\ell + 1}\Bigl[
		(\ell + 1)\Pi_{\dr,\ell+1} - \ell\Pi_{\dr,\ell-1}
	\Bigr] = \alpha_\ell\kddr\Pi_{\dr,\ell}
	\,,
\end{equation}
where the coefficients $\alpha_\ell$ are model-dependent in general. This coupled system can be solved numerically to fully determine the evolution of DM perturbations in the presence of a coupled DR fluid. The DM drag opacity $\kdc$, which appears in the DM density perturbation equations, is generally related through energy conservation to the DR drag opacity $\kddr$. We assume that the microphysical interaction structure is such that these opacities are related by
\begin{equation}
	\label{eq:rate-dm-dr}
	\kdc = \left(
		\frac{4}{3} \frac{\rho_{\mathrm{DR}}}{\rho_{\mathrm{DM}}}
	\right) \kddr
	\equiv S\kddr
	.
\end{equation}
This holds e.g. for Thomson scattering.

The damping scales described above can be observed directly in numerical solutions of these Boltzmann equations when $\kddr$ is normalized to yield a delayed kinetic recoupling.

\section{Kinetic recoupling}
\label{sec:recoupling}
We now consider a different scenario: suppose that the DM and DR baths depart from kinetic equilibrium as usual at $T_\kd$, but then \emph{recouple} later on at $T_\kr$ due to an increase in the microscopic scattering rate, finally decoupling again at $T_\kd'$. In the language of \cref{eq:boltzmann-dm-1,eq:boltzmann-dm-2,eq:boltzmann-dr-1,eq:boltzmann-dr-2}, this amounts to a late-time increase in $\kdc$ such that it again exceeds the conformal Hubble rate $\mathcal H$. In this section, we first introduce mechanisms that can lead to such a thermal history, and then discuss the impact on the evolution of DM perturbations and structure formation.

While our focus in this section is on the formalism rather than on numerics, we show some numerical results produced with the \class{} code \cite{Blas:2011rf} to highlight certain points. We give further details of our numerical procedures in \cref{sec:numerics}.

\subsection{Classes of recoupling scenarios}
\label{sec:realizations}
Several mechanisms can cause the scattering rate to evolve \emph{nonmonotonically,} increasing at late times before decreasing again. These mechanisms generally involve some low-lying scale that becomes relevant as the temperature decreases. As fiducial examples, we consider resonant interactions, phase transitions, or Sommerfeld-enhanced interactions. In particular, none of these effects are typically accounted for in studies based on effective field theory or simplified models. Here we discuss the requirements for a model to exhibit kinetic recoupling.

\begin{figure}
	\centering
	\includegraphics[width=\textwidth]{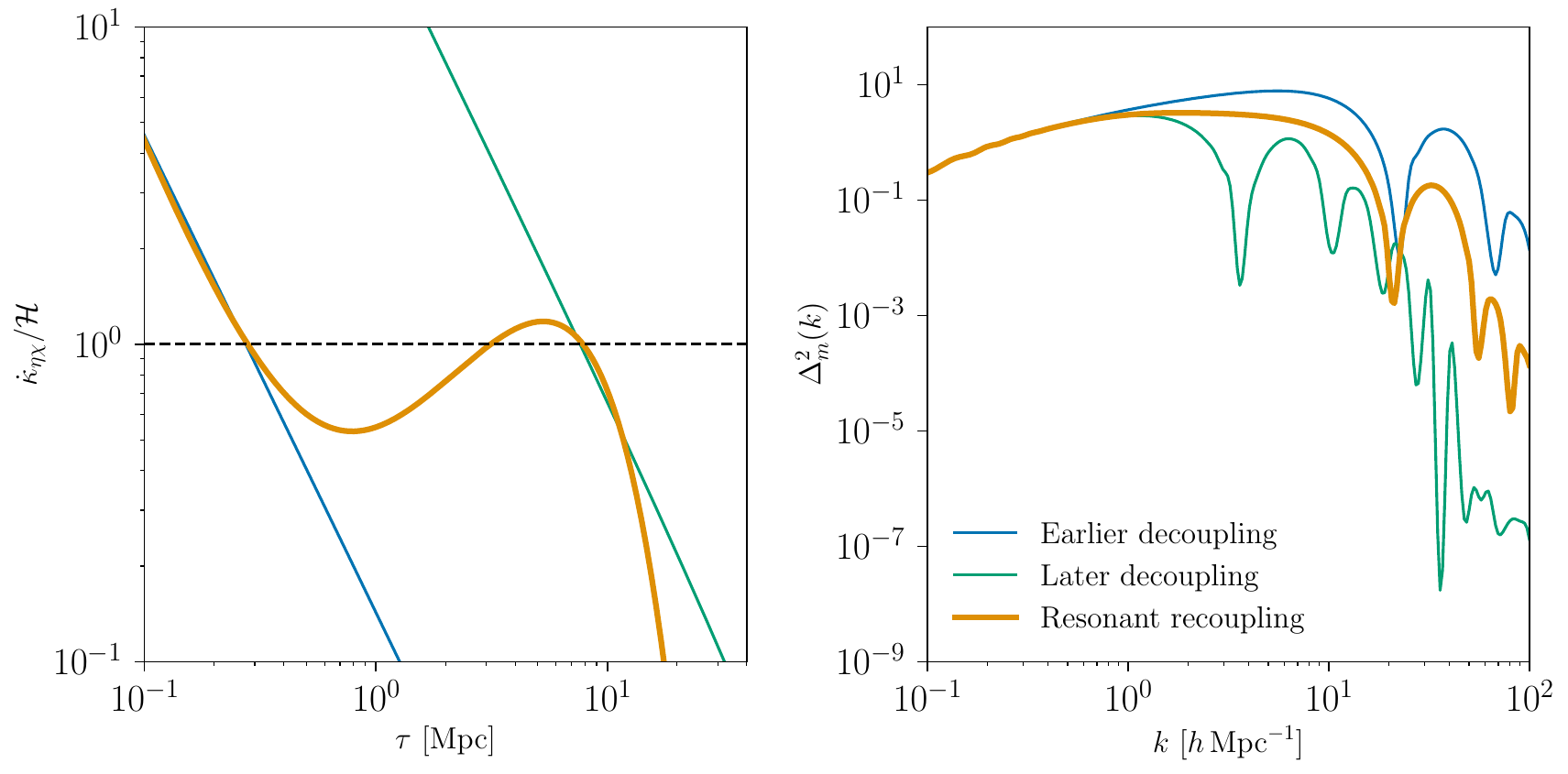}
	\caption{\noindent
	\textbf{Left:} the yellow curve shows the drag opacity as a function of temperature for resonantly-enhanced DM--DR scattering. At late times, $\kddr/\mathcal H$ temporarily deviates from its power-law decrease and exceeds 1, signaling a kinetic recoupling. Here we take $(m_A,~m_\dm,~m_\dr,~g_{\dm\dr}) = (\SI{465}{\electronvolt},~\SI{330}{\electronvolt},~\SI{10}{\milli\electronvolt},~\num{5e-2})$. The blue and green curves show $\kddr$ for single-decoupling scenarios corresponding to the first and second decouplings of the resonant scenario.
	\textbf{Right:} the dimensionless matter power spectrum at $z = 0$ corresponding to each of the scenarios in the left panel.
	}
	\label{fig:resonant-recoupling}
\end{figure}

We first consider the case of a resonant $s$-channel exchange of a real scalar $A$ of mass $m_A$ between fermionic DM $\dm$ and DR $\dr$, i.e., $\dm(p_1)+\dr(k_1)\leftrightarrow A \leftrightarrow \dm(p_2) + \dr(k_2)$. To effect a resonant enhancement at temperatures $T < m_\dm$, we consider $m_\dm = m_A - \delta m$ with $\delta m\ll m_A$. The rate of this process as a function of the DR temperature $T$ is given by \cref{eq:scattering_rate}, where $\omega$ is the energy of the DR in the rest frame of the DM. The center of mass energy is thus $s=(m_\dm+\omega)^2$. The matrix element squared is then given by 
\begin{equation}
	\left|\mathcal{M}\right|^2 =
	\frac{16 g_{\dm\dr}^4\left(s-(m_\dm+m_\dr)^2\right)^2}{
		\left(s-m_A^2\right)^2+m_A^2\Gamma_A^2}
	,
\end{equation}
where $g_{\dm\dr}$ is the $A\dm\dr$ coupling and $\Gamma_A$ is the width of $A$. Since the squared matrix element for the $A \rightarrow \dm\dr$ decay channel scales as $\left|\mathcal M\right|^2\propto g_{\dm\dr}^2$, and the relevant mass scale is $m_A \sim m_\dm \gg m_\dr$, we set $\Gamma_A \approx g_{\dm\dr}^2 m_A$ for simplicity. Lastly, the squared center of mass momentum is given by
\begin{equation}
	k_{\mathrm{cm}}^2  = \frac{1}{4s}
		\left(s^2-2s(m_\dm^2+m_\dr^2)+\left(m_\dm^2-m_\dr^2\right)^2\right).
\end{equation}

At high temperatures, the scattering rate scales as $\Gamma \propto T^4$ since the typical energy exchanged through scattering is of order $T$ while the number density of a relativistic species goes as $T^3$. The Hubble rate in radiation domination scales as $H \propto T^2$. Thus, as in standard scenarios, the ratio $\Gamma(T)/H(T)$ eventually drops too low to maintain efficient momentum transfer, marking kinetic decoupling. When the temperature nears the value of the mass splitting, i.e., $T \approx m_A - m_\dm$, the scattering rate experiences a resonant enhancement. This increases the rate more rapidly than $T^{-2}$, allowing the ratio $\Gamma(T)/H(T)$ to become large. However, as the temperature continues to drop, the resonance vanishes, and the rate again declines faster than $T^2$. This prompts a second, final kinetic decoupling. We illustrate this possibility in  \cref{fig:resonant-recoupling}.

We also considered the possibility of Sommerfeld enhancement re-establishing kinetic equilibrium. Under certain conditions on the DM and mediator masses, Sommerfeld enhancement has been shown to produce \emph{chemical} recoupling, albeit while still in kinetic equilibrium throughout \cite{Feng:2010zp}. However, the onset of the Sommerfeld enhancement requires that the relative velocity of the particles involved is small. One could envision two nonrelativistic populations, one of which is kinetically decoupled, while the other is still in thermal equilibrium. Scattering of these two species could undergo a Sommerfeld enhancement. However, the scattering rate will always be subject to exponential suppression due to the declining number density of the nonrelativistic species, which is stronger than the power-law increase in the Sommerfeld-enhanced scattering cross section. A possibly viable but more involved scenario is that the number density is produced by a large particle-antiparticle asymmetry \cite{Kaplan:2009ag,Petraki:2013wwa}, thereby avoiding the exponential Boltzmann suppression. We leave concrete models that produce kinetic recoupling with Sommerfeld enhancement for future studies.

Kinetic recoupling with a \textit{nonrelativistic} species is also possible through a long-range interaction, e.g. Coulomb scattering, as discussed by \refcite{Kamada:2016qjo}. In that scenario, CHAMPs can fall out of kinetic equilibrium when $T \lesssim m_{e}$ and $e^+ e^-$ annihilation suddenly depletes the abundance of charged particles and thus the scattering rate. After electron-positron annihilation, the dominant contribution to the momentum transfer rate scales as $\kdc/\mathcal H \sim T^{3/2}$ while $H \sim T^2$ during radiation domination, causing $\kdc/\mathcal H$ to increase as $T^{-1/2}$ until recombination. At that point, the rate again falls as the ionization fraction sharply decreases. Depending on the CHAMP mass, this can lead to an initial period of decoupling followed by a recoupling.

In a similar vein, one might consider a recoupling to a radiation bath via a long-range interaction. For instance, if the dark sector consists of a massive particle $\dm$ coupled to a light dark photon $A'$ that kinetically mixes with the Standard Model (SM) photon $\gamma$, resonant transitions between $\gamma$ and $A'$ are possible \cite{Berlin:2022hmt}. In this case, the SM and dark sectors are kept in equilibrium at early times before eventually decoupling. The dark photon receives an effective mass from $\dm$ that, when comparable to the plasma mass of the SM photon, can later produce a resonant enhancement of $\gamma \leftrightarrow A'$ conversions. If this takes place when the SM energy density dominates during the radiation era, then the $A'$ abundance increases significantly. The scattering between $A'$ and $\dm$ is thus enhanced, potentially leading to a kinetic recoupling. However, stringent CMB limits on the abundance of millicharged DM restrict the viability of such a scenario \cite{Dubovsky:2003yn}.

Another class of models involves a phase transition at low temperatures. If the effective scattering rate is increased following the phase transition, this can readily lead to the re-establishment of kinetic equilibrium. Concretely, we consider an interaction Lagrangian of the form
\begin{equation}
	\mathcal{L} \supset g\dm\dr\phi + \lambda\dm\dr\phi\Phi - V(\Phi,\phi)
	,
\end{equation}
where $\phi$ and $\Phi$ are scalar fields interacting via a potential $V(\Phi,\phi)$. Now suppose that the scalar potential is such that $\Phi$ attains a large vacuum expectation value (VEV) $\langle\Phi\rangle = v$ in a phase transition at temperature $T_\eta = T_c$ after the DM has kinetically decoupled, while $\langle\phi\rangle$ remains at zero. For example, if $V(\Phi,\phi) = \alpha_2(T-T_0)\Phi^2 - \alpha_3T\Phi^3+\frac14\alpha_4\Phi^4$ with $\alpha_3\ll\alpha_2,\,\alpha_4$, $\Phi$ acquires a VEV $\langle\Phi\rangle \simeq \sqrt{2(T_0 - T)/\alpha_4}$. Then the Lagrangian acquires an interaction term of the form $\mathcal{L} \supset \left(g + v\lambda\right)\dm\dr\phi$. After symmetry breaking, this term induces a large 3-point interaction between the scalar $\phi$ and the dark sector species $\dm$ and $\dr$.

Explicitly, we write the induced 3-point interaction in the form $\tilde{g}(T)\dm\eta\phi$, where $\tilde{g}(T > T_{c}) = g$ and $\tilde{g}(T<T_{c}) \sim v\lambda \gg g$. In this framework, the thermal scattering rate between dark matter and dark radiation will suddenly spike when $T_\eta\simeq T_{c}$. If we assume the scattering rate to be a power law in temperature, then the opacity scales as
\begin{equation}
	\label{eq:pt-rate}
	\kddr(T) \propto \tilde{g}^2(T)T^{n} =
	T^{n}\begin{cases}
		     g^2          & T > T_{c}, \\
		     v^2\lambda^2 & T < T_{c}. \\
	     \end{cases}
\end{equation}
Clearly, depending on the details of the model, this scenario can lead to $\kddr/\mathcal H > 1$ after symmetry breaking, and thus lead to kinetic recoupling. Since this case is exceptionally simple, we will adopt a form of the phase-transition--modulated rate for our later numerical analysis.

\subsection{Recoupling and structure formation}
Having outlined concrete realizations of the kinetic recoupling scenario, we now discuss the consequences for structure formation.

Physically, a recoupling period can have two effects: (1) it may introduce additional collisional damping for a short time, damping structure on a restricted set of scales; and (2) it may interfere with processes that would otherwise damp structure on particular scales. The first of these is easily understood. For the second, consider the damping of structure by free streaming. The length scale of damping is effectively set by the distance DM particles freely traverse before becoming nonrelativistic. If kinetic recoupling occurs during the free-streaming epoch, the mean free path of DM particles becomes small again, and the free streaming process is interrupted---\emph{reducing} the suppression of the power spectrum. Which of these two effects dominates is model-dependent, and depends on both the timing of the recoupling epoch and the relative importance of free streaming in the absence of the recoupling.

Note that if the first decoupling period is short compared to all other relevant scales, i.e., in the limit $\Delta\tau_\kd \equiv \tau_\kr - \tau_\kd \to 0$, the matter power spectrum should closely match that produced with a single decoupling at $\tau_\kd'$. In this limit, the DM effectively free streams for only a very short time during the first decoupling period, erasing structure on scales much smaller than those that would be affected after the second decoupling. However, even a very brief recoupling period $\Delta\tau_\kr \equiv \tau_\kd' - \tau_\kd$ can have significant, and potentially observable, consequences on the power spectrum. A short period of recoupling that occurs during DM free-streaming is sufficient to scatter a large fraction of the DM particles at least once, reducing the mean free path as above. Moreover, a short period of viscous coupling to the DR bath is enough to significantly affect the phase space distribution of the DM fluid.

Fully assessing the impact of recoupling scenarios on the power spectrum requires numerical solution of the Boltzmann equations, and we will study such numerical solutions in what follows. However, to establish the basic features introduced into the matter power spectrum by a brief recoupling, it is instructive to study the response of the DM density perturbation $\delta_\dm(k,\tau)$ to a \emph{momentary} recoupling, i.e., in the limit $\Delta\tau_\kr \to 0$. This is largely tractable by direct manipulation of the relevant Boltzmann equations, \cref{eq:boltzmann-dm-1,eq:boltzmann-dm-2,eq:boltzmann-dr-1,eq:boltzmann-dr-2,eq:boltzmann-dr-3}.

\begin{figure}
	\centering
	\includegraphics[width=\textwidth]{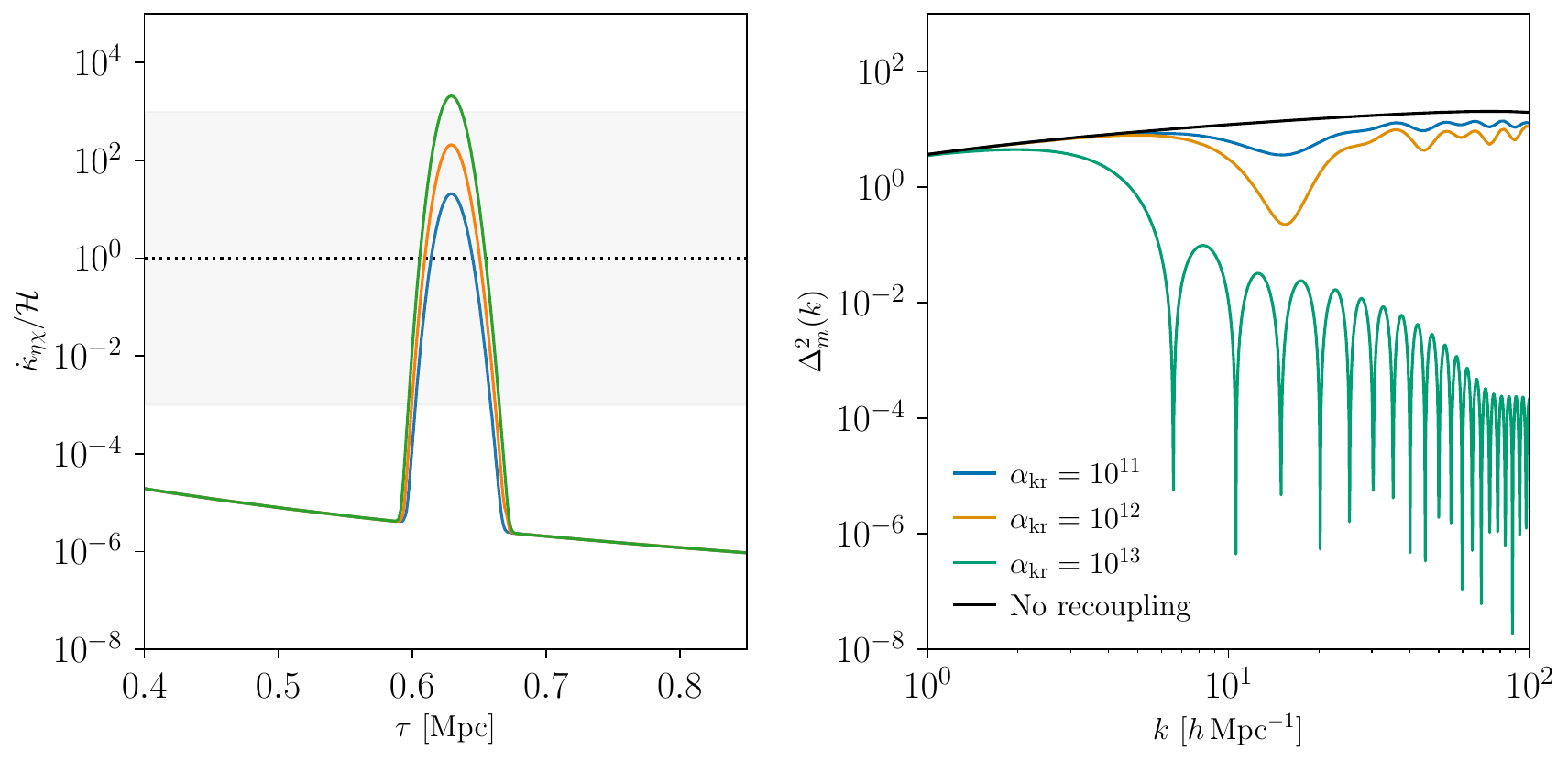}
	\caption{\ignorespaces
	\textbf{Left:} the drag opacity used to simulate a momentary recoupling in \class{} (see \cref{eq:numerical-delta}). Here $\sigma = 0.01$, and three different peak heights $\alpha_\kr$ are shown. The gray band shows the region $\num{e-3} < \kddr/\mathcal H < \num{e3}$, in which \class{} switches between different approximation schemes.
	\textbf{Right:} the dimensionless matter power spectrum at $z = 0$ corresponding to each momentary recoupling.}
	\label{fig:delta-power-spectrum}
\end{figure}

To that end, consider the Boltzmann equations in the case that $\kddr/\mathcal H$ abruptly becomes large. At the time of recoupling, the tight coupling approximation (TCA) becomes valid momentarily. Under the TCA, the anisotropic stress $\sigma_\dr$ and DM sound speed $c_\dm$ are both small and can be neglected, and we have $\delta_\dr \approx \frac43\delta_\dm$. We therefore set
\begin{equation}
	\delta_\dr(\tau) \equiv
	\tfrac43\delta_\dm(\tau)\left[1 + \epsilon(\tau)\right]
	,
\end{equation}
and take $\epsilon(\tau) \ll 1$ during the epoch of recoupling. Moreover, since $\epsilon$ remains fixed during a period of tight coupling, we additionally take $\dot\epsilon(\tau) \ll 1$ during the recoupling. We also write $\dot\epsilon$ in terms of $\dot\delta_\dm$ and $\dot\delta_\dr$, and take $\kdc = S\kddr$. Under these conditions, the Boltzmann equations become
\begin{equation}
	\label{eq:boltzmann-epsilon-1}
	\begin{array}{ll}
		\theta_\dm + \dot\delta_\dm - 3\dot\phi = 0,
		&
		\dot\theta_\dm = k^2\psi - \mathcal H\theta_\dm + 
			(\theta_\dm - \theta_\dr)S\kddr,
		\\
		\theta_\dr + \dot\epsilon\delta_\dm + (1+\epsilon)\dot\delta_\dm
			- 3\dot\phi = 0,
			\qquad\strut
		&
		\dot\theta_\dr = k^2\psi + \tfrac13k^2(1+\epsilon)\delta_\dm +
			(\theta_\dr - \theta_\dm)\kddr
		.
	\end{array}
\end{equation}

Before expanding in $\epsilon$ and $\dot\epsilon$, some care is required due to the large scattering rate during the recoupling period. Nominally, the DM--DR slip $\Delta\theta \equiv \theta_\dr - \theta_\dm$ should also be $\mathcal O(\epsilon)$ during a period of tight coupling, so we write $\Delta\theta(\tau) \equiv \Omega(\tau)\epsilon(\tau)$. On the other hand, the drag opacity becomes large during this epoch, and the product $\kddr\times\Delta\theta$ is not small---indeed, this is exactly what drives the slip to zero. Thus we write $\kddr(\tau) \equiv \gamma(\tau)/\epsilon(\tau)$. Making these replacements, \cref{eq:boltzmann-epsilon-1} becomes
\begin{equation}
	\label{eq:boltzmann-epsilon-2}
	\begin{array}{ll}
		\theta_\dm + \dot\delta_\dm - 3\dot\phi = 0,
		&
		\dot\theta_\dm = k^2\psi - \mathcal H\theta_\dm -
			\Omega S\gamma,
		\\
		\theta_\dm + \epsilon\Omega + \dot\epsilon\delta_\dm
			+ (1+\epsilon)\dot\delta_\dm - 3\dot\phi = 0,
			\quad\strut
		&
		\dot\theta_\dm = k^2\psi + \tfrac13k^2(1+\epsilon)\delta_\dm +
			\Omega\gamma - \tfrac13\partial_\tau(\epsilon\Omega)
		.
	\end{array}
\end{equation}
To zeroth order in $\epsilon$ and $\dot\epsilon$, this system reduces to three independent equations:
\begin{equation}
	\label{eq:boltzmann-epsilon-3}
	\theta_\dm + \dot\delta_\dm - 3\dot\phi = 0,
	\quad
	\dot\theta_\dm = k^2\psi - \mathcal H\theta_\dm - \Omega S\gamma,
	\quad
	- \mathcal H\theta_\dm =
		\tfrac13k^2\delta_\dm + (1+S)\Omega\gamma
	.
\end{equation}
In this form, a few conclusions are manifest. First, a momentary recoupling can cause discontinuities in the velocity perturbations $\theta_i$, and therefore in $\dot\delta_i$, but not in $\delta_i$, $\phi$, or $\psi$. This is intuitive: if the rate of momentum exchange is suddenly made large, the velocities change immediately, but the densities lag. Second, when $\kddr$ becomes large, the response of $\theta_\dm$ is strongly scale-dependent. In particular, after a short decoupled period $\Delta\tau_\kd$, $\left|\Omega(k)\right|$ grows strongly with $k$, since free streaming erases small-scale perturbations only in the radiation bath. Since $\gamma(\tau)$ is independent of $k$, this suggests that $\theta_\dm$ and $\dot\theta_\dm$ will exhibit much more dramatic changes on small scales at the moment of recoupling. But third, at very small scales, the $k^2\psi$ and $k^2\delta$ terms will dominate, minimizing the effect. Finally, such dramatic shifts will tend to drive $\Omega$ away from being large, which will in turn drive $\dot\theta$ away from being large.

Thus, we can expect $\theta_\dm$ and $\theta_\dr$ to suddenly snap together at the moment of recoupling, then evolve slowly together during the recoupled period, and then separate again after the second decoupling. We expect this effect to be significant at intermediate scales, and less significant on either very large or very small scales. Note that there are \emph{two} length scales associated with a momentary recoupling: one is the horizon scale at recoupling, and one is the duration of the first decoupled epoch, $\Delta\tau_\kd$. The latter scale is crucial because it reflects the time during which the DM perturbations and the DR perturbations can evolve independently. In this decoupled period, the DM perturbations grow and the DR perturbations are suppressed. Thus, $\Delta\tau$ directly controls the size of $\Delta\theta$ at $\tau_\kr$, and in turn, this controls the size of the effect when $\Delta\theta$ is driven to zero by the recoupling.

Based on this picture, we can go one step further and use the system of \cref{eq:boltzmann-epsilon-1} to estimate $\theta_\dm$ immediately after the recoupling. At this time, we know that we must have $\theta_\dm = \theta_\dr$, but since both of these perturbations are evolving, we have yet to determine where they meet. To that end, we linearize, taking $\theta_i(\tau) \approx \theta_i(\tau_0) + \dot\theta_i(\tau_0)\,(\tau - \tau_0)$. The rapid evolution of $\theta_i$ stops when $\theta_\dm = \theta_\dr$. Solving for the time at which the linear approximants of $\theta_\dm$ and $\theta_\dr$ coincide, we find
\begin{equation}
	\label{eq:theta-estimate}
	\theta_\dm(\tau_\kr) = \theta_\dr(\tau_\kr) \simeq
	\left.\frac{
		k^2\delta_\dr\theta_\dm-4k^2\psi\,\Delta\theta
		+4\mathcal H\theta_\dm\theta_\dr+4\,\Delta\theta
		\,(\theta_\dm+S\theta_\dr)\kddr
	}{
		k^2\delta_\dr+4\mathcal H\theta_\dm+4(1+S)\,\Delta\theta\,\kddr
	}\right|_{\tau=\tau_0}
	,
\end{equation}
where $\tau_0$ is chosen to be just before the recoupling time $\tau_\kr$.

Let us pause to interpret this result. The equation directly shows that the recoupling interrupts the normal evolution of the DM velocity perturbation, suddenly averaging it with the DR velocity perturbation. This translates to a sudden shift in the derivative of the DM density perturbation, so the momentary recoupling effectively gives $\delta_\dm$ a sudden \textit{kick} in one direction or the other. The size and direction of this kick are scale-dependent. To be clear, then, \cref{eq:theta-estimate} does not eliminate the need to solve for the evolution of perturbations as usual. Rather, it provides an quantitative estimate of the effect of the kick in terms of the conditions under which recoupling takes place. In principle, this could be taken as a prescription for implementing an instantaneous recoupling in a Boltzmann solver without designing an interaction rate that effects such a recoupling.

\begin{figure}\centering
	\includegraphics[width=\textwidth]{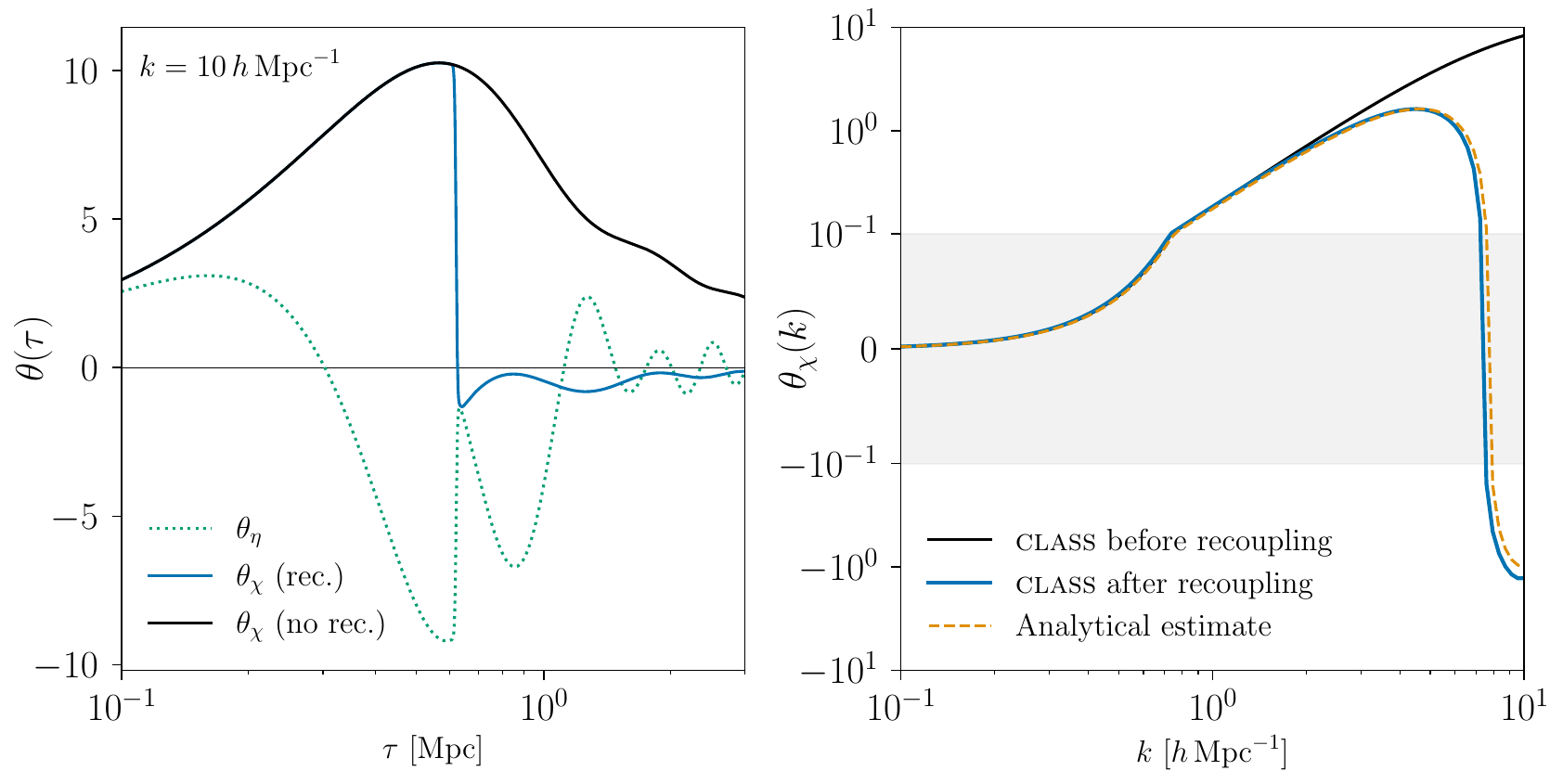}
	\caption{\ignorespaces
		Numerical evolution of velocity perturbations during a momentary recoupling, as in \cref{eq:numerical-delta} and \cref{fig:delta-power-spectrum}.
		\textbf{Left:} the evolution of $\theta_\dm$ and $\theta_\dr$ with $\tau$ at a fixed wavenumber $k = 10\hMpc$. The black curve shows the evolution of $\theta_\dm$ in the absence of a recoupling. The DM and DR velocity perturbations immediately snap together at the moment of recoupling, and thereafter separate once again.
		\textbf{Right:} $\theta_\dm$ immediately before and after recoupling as a function of $k$. The black curve shows $\theta_\dm(\tau_0)$ in the sense of \cref{eq:theta-estimate}, with $\tau_0$ chosen just before $\tau_\kr$. The blue curve shows the value attained just after recoupling, when $\theta_\dm \approx \theta_\dr$. The dashed yellow curve shows the analytical estimate of \cref{eq:theta-estimate}. The vertical axis is linear within the shaded region.
	}
	\label{fig:delta-test}
\end{figure}

For our purposes, however, this is most useful as a benchmark to validate our intuition against a numerical solver. To that end, we verify that \cref{eq:theta-estimate} approximately holds in comparison with numerical integration of the Boltzmann equations using \class{} (see \cref{sec:numerics} for details). We implement the sharp recoupling as a narrow Gaussian contribution\footnote{We model the recoupling as a smooth contribution to the opacity in order to avoid numerical artifacts.} to $\kddr$ such that the total opacity is given by
\begin{equation}
	\label{eq:numerical-delta}
    \kddr \propto
    	(1 + \alpha_\kr \times e^{-\Delta^2 / \sigma^2}) \times T^{n},
\end{equation}
where $n$ is the power law index of the opacity in temperature away from the recoupling, $\sigma$ controls the width of the Gaussian peak, and $\Delta \equiv T_\dr - T_{\dr,\kr}$ determines the dark radiation temperature at which the recoupling takes place. (Recall that the dark radiation temperature is in general different from the Standard Model bath temperature.) The form of $\kddr$ is shown in \cref{fig:delta-power-spectrum} for several values of $\alpha_\kr$, together with the matter power spectrum for each case.

\begin{figure}\centering
	\includegraphics[width=\textwidth]{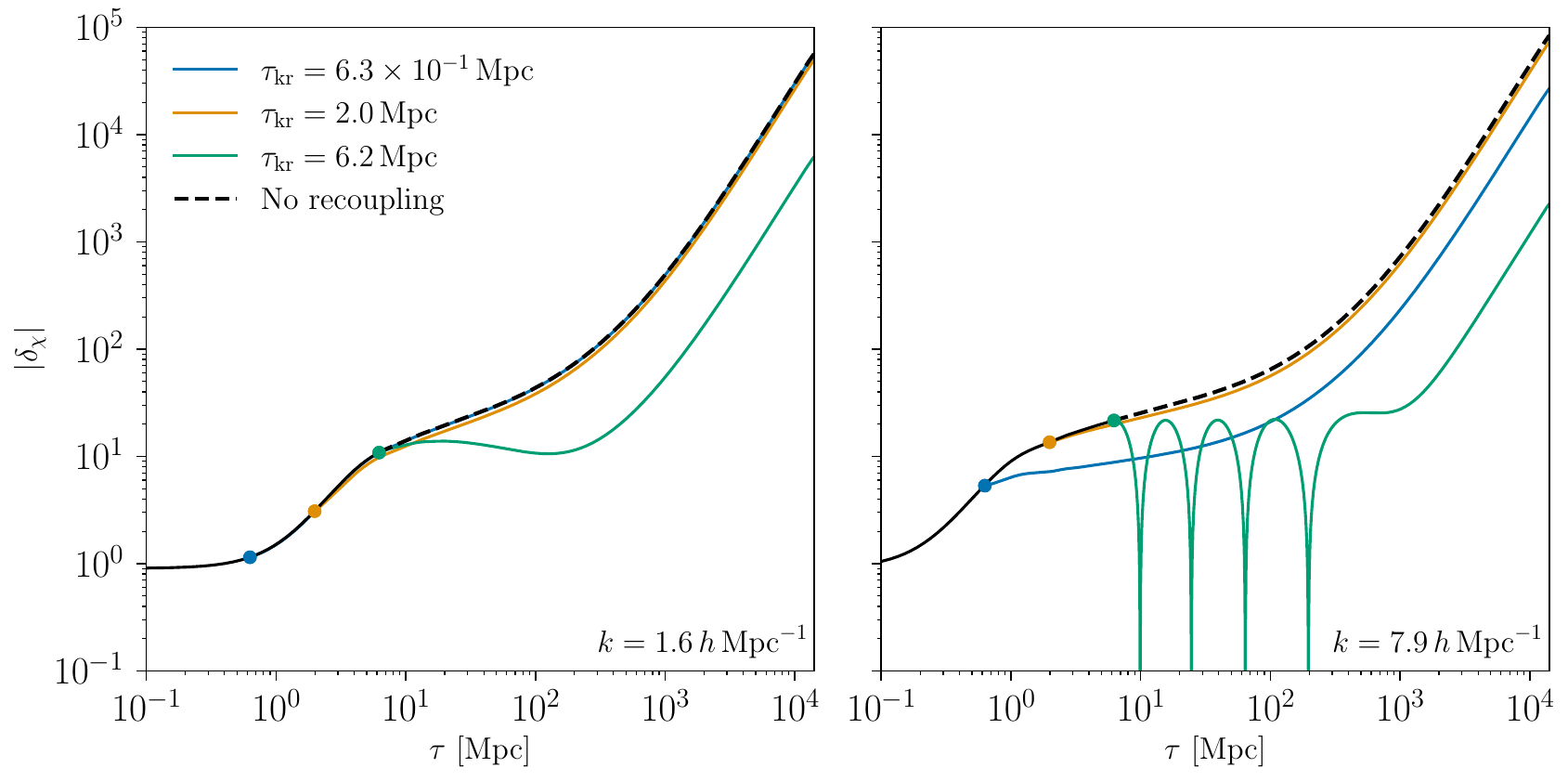}
	\caption{\ignorespaces
		evolution of $\delta_\dm$ as a function of conformal time with a momentary recoupling for three different recoupling times $\tau_\kr$, each indicated by a solid dot, and for two different wavenumbers in the two panels. The dashed black curve shows $\delta_\dm(\tau)$ with no kinetic recoupling. The size of the kick to $\delta_\dm$ is manifestly strongly dependent on both $\tau_\kr$ and $k$, but generally acts to delay the formation of structure. In particular, in the right panel, the latest recoupling time shown pushes $\delta_\dm$ into an oscillatory period, substantially delaying the growth of structure in that mode.
	}
	\label{fig:recoupling-effect}
\end{figure}

With this implementation of a momentary recoupling, the estimate for $\theta_\dm$ in \cref{eq:theta-estimate} is validated explicitly in \cref{fig:delta-test}.\footnote{We caution that the analytical estimate is extremely crude, and should serve only as a heuristic for the parametrics of recoupling. The striking agreement in \cref{fig:delta-test} should not be taken as evidence that it is robust in general.} The numerical result also verifies the heuristic statements above regarding the effect of the recoupling on the perturbations themselves. The behavior of $\theta$ is as expected from the TCA: the DM and DR velocity perturbations, $\theta_\dm$ and $\theta_\dr$, instantly snap to a common value. After the recoupling, $\theta_\dm$ and $\theta_\dr$ are once again free to vary independently, but retain a memory of the period of recoupling. Accordingly, $\delta_\dm$ undergoes a discontinuity in its derivative, or a \textit{kick}, as we described it above. The effect of this kick is shown directly in \cref{fig:recoupling-effect}. A small kick essentially shifts the time evolution of $\delta_\dm$, slightly delaying its ordinary evolution. A large kick, on the other hand, can push $\delta_\dm$ into the oscillatory regime, substantially delaying the growth of structure in the corresponding mode.

\Cref{fig:resonant-recoupling,fig:delta-power-spectrum} also show the impact of a momentary kinetic recoupling on the matter power spectrum. As might be expected, a stronger, longer recoupling results in a stronger suppression of small-scale structure, and produces a cutoff in the spectrum at larger scales. However, interpretation of these power spectra is subtle: in the cases shown here, the amplitude of the recoupling is intrinsically linked to its duration. In the following section, we elaborate on our numerical procedures, and use further numerical results to elucidate the qualitative behavior of the matter power spectrum following a recoupling.

\section{Numerical analysis}
\label{sec:numerics}
The analysis of the previous section is ultimately limited by the complexity of the differential equations underlying the evolution of DM perturbations. In this section, we undertake the numerical study of kinetic recoupling using the Boltzmann code \class{} \cite{Blas:2011rf}, and we connect various recoupling scenarios to the small-scale matter power spectrum.

\subsection{Implementation in \class{}}
The ETHOS framework naturally accommodates DM--DR interactions, and this formalism has been implemented into the \class{} code by the authors of \refcite{Archidiacono:2017slj}. The principle behind ETHOS is that the DM microphysics can be mapped onto some set of low-energy parameters, enabling the systematic study of observables in this effective parametrization rather than in any specific model. In practice, this means that the scattering rate of DR off of DM is generically modeled as a power law in temperature (or equivalently redshift), as
\begin{equation}
    \kddr =
    -\Omega_{\mathrm{DM}}h^2g_{\dm\dr}\left(\frac{1 + z}{10^7}\right)^n,
\end{equation}
where $g_{\dm\dr}$ is a coupling constant and $n$ is the power law index governing the temperature dependence. This form of the opacity cannot accommodate a kinetic recoupling: such scenarios do not meet the hypotheses of the ETHOS framework. However, the modifications made to \class{} to implement generic DM--DR opacities within ETHOS can be extended to solve the Boltzmann equations with any temperature dependence in $\kddr$, as long as the input $\kddr$ is numerically well behaved. Thus, we implement kinetic recoupling scenarios in \class{} by directly modifying $\kddr$ to the forms corresponding to our recoupling scenarios.

In particular, we implement the phase transition scenario of \cref{eq:pt-rate} in \class{} by modifying the rate as follows: we introduce two new parameters $T_\kr$ and $A_\kr$ that control the temperature and amplitude of the recoupling. The opacity is then treated as a piecewise function
\begin{equation}
	\label{eq:rateDRDM}
	\kddr =
	\begin{cases}
     	-\Omega_{\mathrm{DM}}h^2 g_{\dm\dr} \left(
      		\frac{1 + z}{10^7}\right)^n \quad & \, T > T_\kr, \\
     	-\Omega_{\mathrm{DM}}h^2 g_{\dm\dr} \left(
      		\frac{1 + z}{10^7}\right)^n (1 + A_\kr) \quad & \, T \leq T_\kr.
     \end{cases}
\end{equation}
In order to achieve the required numerical stability, we also increase the sampling of thermodynamic quantities by a factor of $100$ in order to capture the rapid change in behavior at $T_\kr$. (Empirically, further refining the sampling does not alter the output.) The set of parameters shared by all phase transition scenarios implemented in \class{} is given in \cref{tab:class-params}.

\begin{table}[ht]
\centering
    \begin{tabular}{@{}lccl@{}}
    \toprule
    Parameter                & Value  & \strut~~\strut & Description     \\ 
    \midrule
    \texttt{m\_idm}          & $10^3$ & & DM mass [\SI{}{\electronvolt}] \\
    \texttt{a\_idm\_dr}      & 10     & & $g_{\dm\dr}$ (DM--DR coupling) \\
    \texttt{xi\_idr}         & 0.3    & &
    	Temperature ratio $\xi = T_\dr/T_\gamma   $              \\
    \texttt{nindex\_idm\_dr} & 2      & &
    	Power law index of opacity temperature dependence                \\
    \texttt{stat\_f\_idr}    & 0.875  & &
    	Statistical factor ($7/8$ for fermionic DR)                      \\
    \bottomrule
    \end{tabular}
    \caption{\class{} parameters used for the numerical implementation of all phase transition scenarios.}
    \label{tab:class-params} 
\end{table}

\subsection{Signatures in the matter power spectrum}
\label{sec:signatures}
We now consider the features of the matter power spectrum associated with the properties of a kinetic recoupling. We have already treated the simplest cases in \cref{sec:recoupling}. In \cref{fig:resonant-recoupling,fig:delta-power-spectrum}, we observed nontrivial behavior in the matter power spectrum for brief recouplings, and at the broadest level, it is easy to understand why small-scale structure is suppressed by these recouplings: the recoupling delivers a scale-dependent kick to $\delta_\dm$, delaying the formation of structure. This scale dependence is controlled in part by the horizon scale at recoupling (and thus $\tau_\kd)$, and in part by the duration of the first kinetically-decoupled epoch, $\Delta\tau_\kd = \tau_\kr - \tau_\kd$. For a realistic kinetic recoupling, which is not instantaneous, a \emph{third} scale becomes relevant: the duration of the recoupled epoch, $\Delta\tau_\kr$.

\begin{figure}
	\centering
	\includegraphics[width=\textwidth]{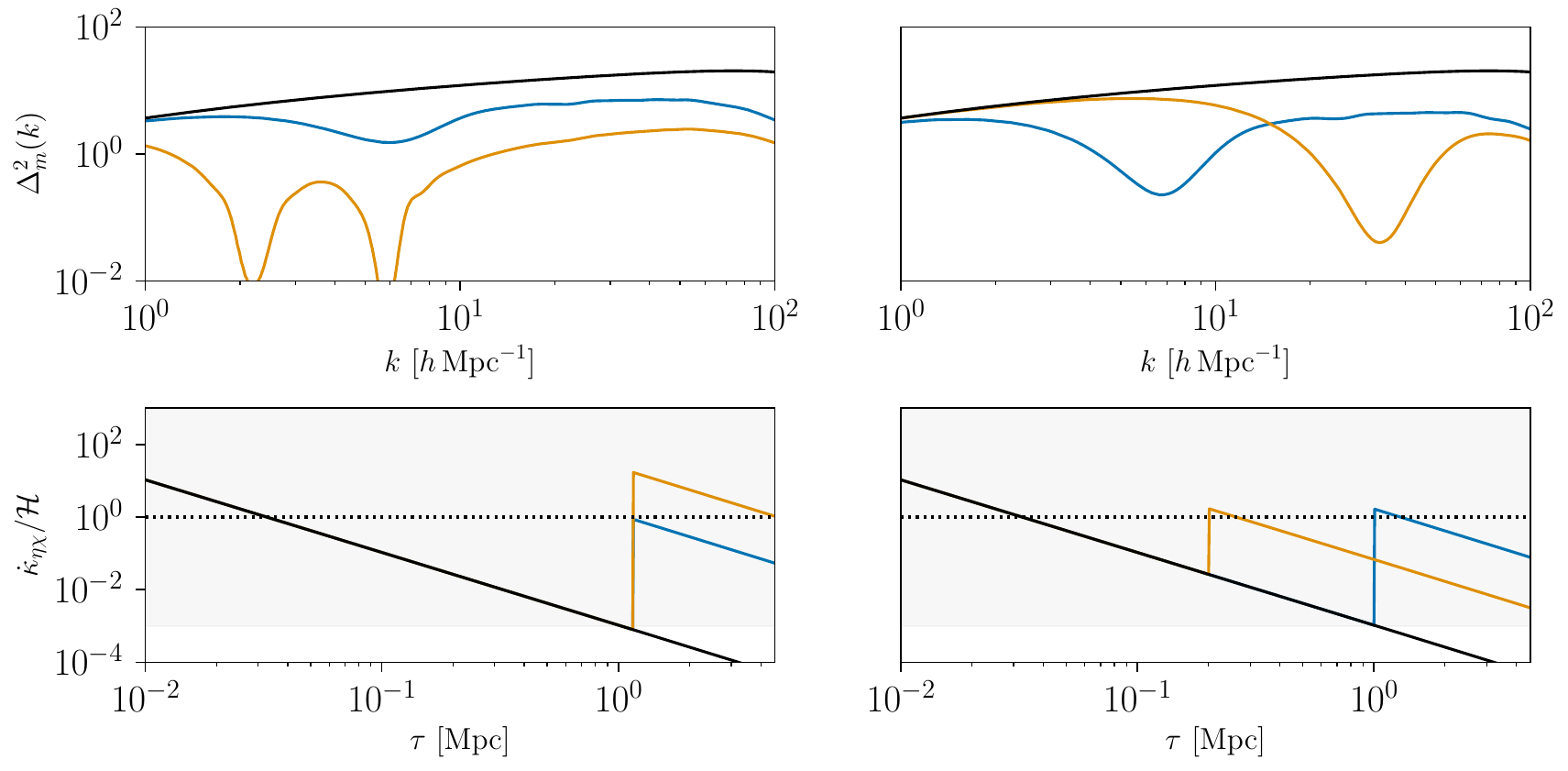}
	\caption{Effects on the dimensionless matter power spectrum for different realizations of the phase transition scenario. The bottom panels show the opacity for different choices of $A_\kr$ and $\tau_\kr$, and the top panels show the corresponding dimensionless matter power spectra. The left panels effectively vary the duration of the recoupling, $\Delta\tau_\kr$, while holding the recoupling time $\tau_\kr$ fixed. The right panels vary the time of recoupling, $\tau_\kr$, while holding $\Delta\tau_\kr$ and $\tau_\kd$ fixed.}
	\label{fig:pt-variations}
\end{figure}

To study the scale dependence in more detail, we must have a toy model in which these scales can be manipulated individually. The scenarios already considered numerically in \cref{sec:recoupling} are inconvenient for this purpose: increasing the Gaussian peak height for a momentary recoupling also increases its duration. It is simpler to work with the phase transition scenario as parametrized in \cref{eq:rateDRDM}. Here, $T_\kr$ and $A_\kr$ effectively control $\tau_\kr$ and $\Delta\tau_\kr$.

This is shown explicitly in the bottom panels of \cref{fig:pt-variations}, and the corresponding power spectra are shown in the top panels. From the right panels, which vary $\tau_\kr$, it is clear that for a momentary recoupling, the horizon size at $\tau_\kr$ determines the location of the inverse peak in the power spectrum. In the left panels, the extended duration of the recoupling translates to the range of wavenumbers that are subject to severe suppression. While the power spectrum is suppressed to some extent at all higher wavenumbers, our previous observation regarding the scale dependence holds: the suppression is \emph{less} significant at very small scales than would be expected for a comparable late decoupling. This means that \emph{the time and duration of a kinetic recoupling are both imprinted on the matter power spectrum.}

We thus consider the prospects for discovering a kinetic recoupling via the matter power spectrum and distinguishing it from other thermal histories. To that end, we first consider the available observational probes of the power spectrum.

The matter power spectrum is  constrained by a variety of cosmological observations as a function of wavenumber $k$. For $10^{-4}\hMpc\lesssim k\lesssim 1\hMpc$, CMB observations \cite{Planck:2018vyg} and large-scale galaxy surveys such as BOSS \cite{Gil-Marin:2015sqa} and DES \cite{DES:2017myr} provide the most stringent constraints. At increasingly smaller scales, $1\hMpc\lesssim k\lesssim 10\hMpc$, Lyman-$\alpha$ forest data and weak lensing provide the best measurements of the matter power spectrum \cite{SDSS:2017bih,DES:2017qwj}. However, larger wavenumbers, and thus smaller scales, remain largely unconstrained observationally.

A highly anticipated possibility is that upcoming \SI{21}{\centi\meter} observations will provide a new probe of the small-scale matter power spectrum. Since this wavelength corresponds to the hyperfine transition of neutral hydrogen, the \SI{21}{\centi\meter} sky measures both resonant absorption of CMB photons and the heating of the intergalactic medium caused by X-ray emissions from the first stars. With future measurements of the global \SI{21}{\centi\meter} signal and its fluctuations, \refcite{Munoz:2019hjh} argues that the EDGES \cite{Leo:2019gwh} and HERA \cite{DeBoer:2016tnn} experiments will provide precise measurements of the matter power spectrum for wavenumbers much larger than $1\hMpc$. Specifically, EDGES is anticipated to measure the matter power spectrum in the range of $40\textnormal{--}80\hMpc$ to 30\% accuracy, while HERA is anticipated to achieve astonishing 1.2\% accuracy for wavenumbers $3\hMpc\lesssim k\lesssim 40\hMpc$, 16\% for $40\hMpc\lesssim k\lesssim 60\hMpc$, and 30\% for $60\hMpc\lesssim k\lesssim 80\hMpc$. Whether or not such accuracy will be reached by these particular experiments is still a matter of great controversy. Still, given these promising possibilities, we assume here that the projections of \refcite{Munoz:2019hjh} will be exactly realized.

\begin{figure}\centering
	\includegraphics[width=\textwidth]{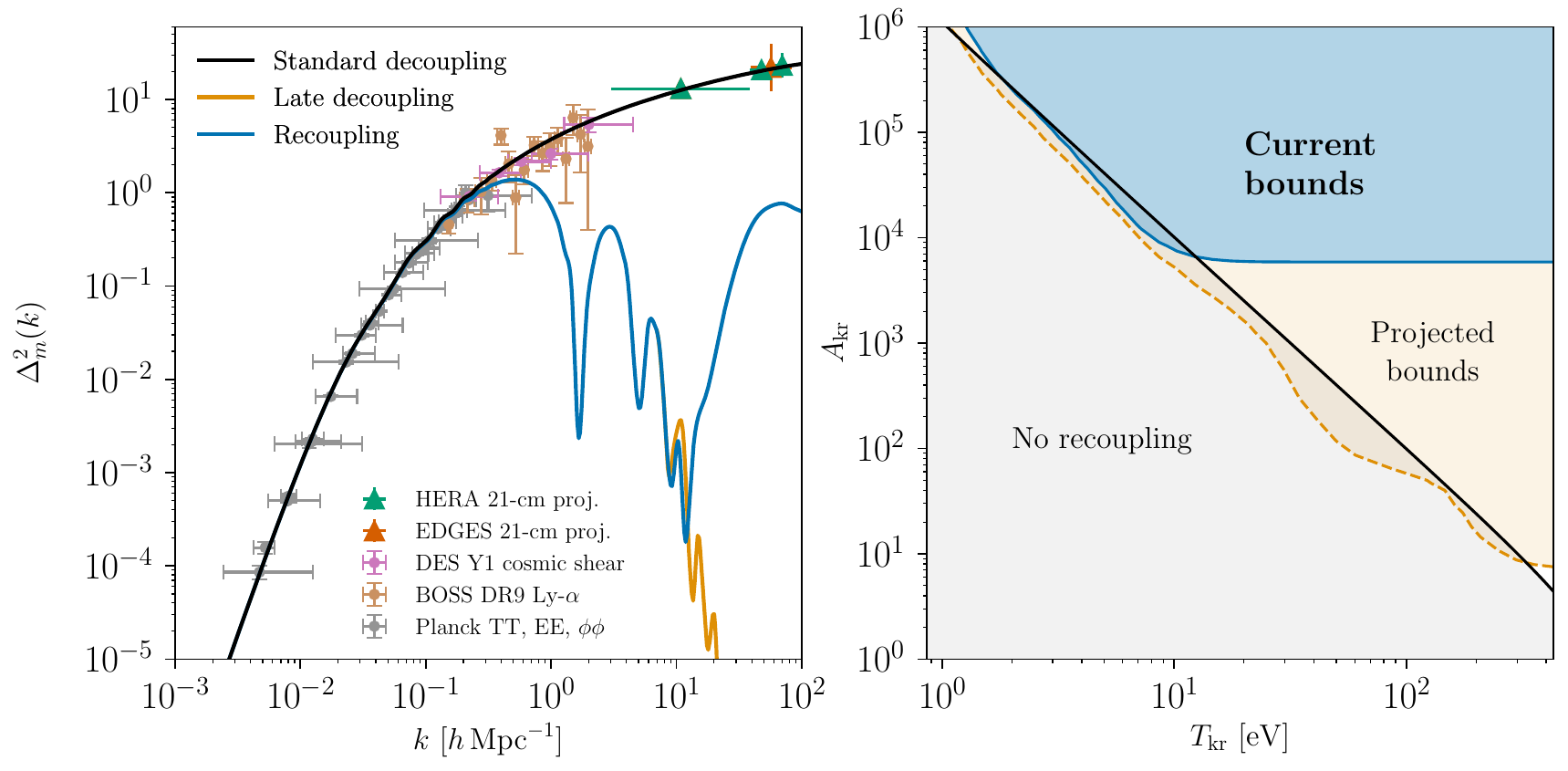}
	\caption{\ignorespaces
	\textbf{Left:} the dimensionless matter power spectrum $\Delta_m^2(k) = P(k) \frac{k^3}{2 \pi^2}$ for our benchmark parameter values. We show the kinetic recoupling scenario as well as the best-fit single-decoupling scenario (see text). Current measurements are shown including CMB data from Planck \cite{Planck:2018vyg}, Lyman-$\alpha$ forest data from BOSS DR9 \cite{Gil-Marin:2015sqa,SDSS:2017bih}, and cosmic shear measurements from DES Y1 \cite{DES:2017myr,DES:2017qwj}. We also show projected measurements from \SI{21}{\centi\meter} data from EDGES \cite{Leo:2019gwh} and HERA \cite{DeBoer:2016tnn}, as derived by \refcite{Munoz:2019hjh}.
	\textbf{Right:} parameter space for a recoupling of the form of \cref{eq:rateDRDM} with the values in \cref{tab:class-params}. The blue region is ruled out at the 95\% confidence level with current measurements. The yellow region shows projected constraints should data from EDGES and HERA match the standard scenario (black curve in the left panel). In the gray region, $A_\kr$ is too small to produce a full recoupling in the sense of $\kddr/\mathcal H>1$.}
	\label{fig:recoupling-spectrum-constraints}
\end{figure}

Given these conditions, we examine a benchmark scenario corresponding to a recoupling at $T_\kr = \SI{114}{\electronvolt}$ with $A_\kr = \num{4.5e4}$. The corresponding power spectrum is shown in the left panel of \cref{fig:recoupling-spectrum-constraints}. These parameter values are chosen for demonstration purposes: they sit slightly outside the allowed parameter space, and represent a scenario that would be particularly amenable to discovery and identification. \Cref{fig:recoupling-spectrum-constraints} also shows the power spectrum corresponding to a late kinetic decoupling. The decoupling time for this curve is chosen to give the best possible least-squares fit to the recoupling curve. That is, the late-decoupling curve represents the closest mimic of kinetic recoupling that could be obtained with a delayed single decoupling. The two curves are notably identical at sufficiently large scales, but as we have noted, the suppression of the power spectrum in the recoupling scenario is relaxed at yet smaller scales. Thus, the two scenarios are manifestly distinct. Under our present hypotheses, they could be readily distinguished by upcoming HERA and EDGES observations.

Current and future power spectrum measurements place constraints on the parameter space of kinetic recoupling, shown in the right panel of \cref{fig:recoupling-spectrum-constraints}. Strictly speaking, these constraints apply to the phase transition scenario with the parameter values in \cref{tab:class-params}, but $A_\kr$ and $T_\kr$ together control the timing and duration of the recoupling epoch. The region below the black line does not produce a full recoupling, since $A_\kr$ is too small to yield $\kddr/\mathcal H>1$. Current measurements from Planck, BOSS, and DES rule out even a brief kinetic recoupling at temperatures $T \lesssim \SI{10}{\electronvolt}$. Future measurements from HERA and EDGES will constrain momentary recouplings at temperatures up to \SI{400}{\electronvolt}. Note that the constraints shown in \cref{fig:recoupling-spectrum-constraints} do not extend to arbitrarily high temperatures. In the context of our scenario, we must have $T_\kr < T_\kd$, and the latter is $\mathcal O(\SI{}{\kilo\electronvolt})$ for our benchmark parameters.

\section{Discussion and conclusions}
\label{sec:discussion}
In this work, we have considered the possibility that the cosmological dark matter re-enters kinetic equilibrium with a relativistic thermal bath after an initial decoupling. We anticipated that such a new kinetic equilibrium phase would lead to additional suppression in the matter power spectrum, and outlined a heuristic description of the effects. We then surveyed a few plausible scenarios where such a kinetic recoupling might occur: a resonantly-enhanced scattering process, a phase transition altering the coupling between the dark matter and the radiation bath, or a Sommerfeld enhancement effect at low velocities.

While these scenarios provide a concrete basis for the rest of our discussion, this list is not exhaustive. One other possibility is a dark sector with a confining species where the confined bound state possesses a large scattering cross section with a relativistic bath, suppressed in the unconfined phase. Another possibility is that of a modified cosmology such that the Hubble rate undergoes a large enhancement, driving a first kinetic decoupling, followed by a second one when the Hubble rate proceeds to drop. Lastly, long-range dark-sector interactions lead to plasma effects that could also produce a large late-time scattering rate.

We implemented numerical examples of kinetic recoupling in \class{}, utilizing the ETHOS framework \cite{Vogelsberger:2015gpr}. We computed power spectra corresponding to several recoupling scenarios. We found that the matter power spectrum undergoes significant suppression at small scales, with the largest effects on scales of order the horizon size at recoupling. An extended period of recoupling therefore corresponds to suppression on an extended range of scales, but recoupling generically features less suppression on very small scales than would be anticipated in other scenarios.

Finally, we studied observational prospects, including the possibility of distinguishing between kinetic recoupling and other mechanisms known to suppress power at small scales. We found that kinetic recoupling predicts unique nontrivial effects in the power spectrum at sufficiently small scales, as shown in \cref{fig:recoupling-spectrum-constraints}. If future \SI{21}{\centi\meter} surveys succeed in probing the power spectrum on scales on the order of $k\sim 100\hMpc$, the discovery of a kinetic recoupling in cosmological data is a realistic prospect.

\acknowledgments

We gratefully acknowledge valuable conversations with
Asher Berlin,
Innes Bigaran,
Kim Boddy,
Nick Gnedin,
Roni Harnik,
Alejandro Ibarra,
Yoni Kahn,
David Kaiser,
Hiren Patel,
and Jessie Shelton.
The work of B.V.L. was supported in part by the Josephine de Karman Fellowship Trust, by the Sandwich Fellowship Program at Hebrew University, and by the MIT Pappalardo Fellowship. The work of N.S. was supported in part by the NSF Graduate Research Fellowship Program under grant No. DGE-1842400. The work of B.V.L. and S.P. was supported in part by DOE grant No. DE-SC0010107.

\bibliographystyle{JHEP}
\bibliography{main}

\end{document}